\documentclass[aps, showpacs, nofootinbib, preprintnumbers, twocolumn]{revtex4}
\usepackage{amssymb,amsmath,microtype}
\usepackage{graphicx}
\usepackage[normalem]{ulem}

\usepackage{color}

\newcommand{\beq}{\begin {equation}}
\newcommand{\eeq}{\end   {equation}}
\newcommand{\bea}{\begin {eqnarray}}
\newcommand{\eea}{\end   {eqnarray}}
\newcommand{\baa}{\begin {array}   }
\newcommand{\eaa}{\end   {array}   }
\newcommand{\bit}{\begin {itemize} }
\newcommand{\eit}{\end   {itemize} }
\newcommand{\be }{\begin {equation}}
\newcommand{\ee }{\end   {equation}}
\newcommand{\nn }{\nonumber        }

\begin{document}

\preprint{ACFI-T17-06}

\title{Explore Inert Dark Matter Blind Spots with Gravitational Wave Signatures}

\author{Fa Peng Huang}
\email{huangfp@ihep.ac.cn}
\affiliation{Theoretical Physics Division, Institute of High Energy Physics, Chinese Academy of Sciences, P.O.Box 918-4, Beijing 100049, P.R.China}

\author{Jiang-Hao Yu}
\email{jhyu@physics.umass.edu}
\affiliation{Amherst Center for Fundamental Interactions, Department of Physics,
University of Massachusetts Amherst, Amherst, MA 01002, USA}

\begin{abstract}
Motivated by the absence of dark matter signals in direct detection experiments  (such as the recent LUX and PandaX-II experiments) and
the  discovery of gravitational waves (GWs) at aLIGO,
we discuss the possibility to explore a generic classes of scalar dark matter models using the complementary searches via phase transition GWs  and the future lepton collider signatures.
We focus on the inert scalar multiplet dark matter models and the mixed inert scalar dark matter models, which could undergo a strong first-order phase transitions during the evolution of the early universe, and might produce detectable phase transition GW signals at future GW experiments, such as eLISA, DECIGO and BBO.
We find that the future GW signature, together with circular electron-positron collider, could further explore the
model's {\it blind spot} parameter region, at which the dark matter-Higgs coupling is identically zero,
thus avoiding the dark matter spin-independent direct detection constraints. 

\end{abstract}


\maketitle

\section{Introduction}
A brand new door has been opened to study the fundamental physics and the particle cosmology by the gravitational waves (GWs)
after the discovery of the GWs by Advanced Laser Interferometer Gravitational Wave Observatory (aLIGO)~\cite{Abbott:2016blz}.
One active research field is the idea of probing the new physics through phase transition GWs, where a strong first-order phase transition (FOPT)
is induced by the new physics models and can produce detectable GW signals from three mechanisms: collisions of expanding
bubbles walls, magnetohydrodynamic turbulence of bubbles and sound waves in the hot plasma of the early universe~\cite{Witten:1984rs, Hogan:1984hx, Turner:1990rc,Kamionkowski:1993fg,Hindmarsh:2013xza,Hindmarsh:2015qta,Kosowsky:2001xp,Caprini:2009yp}.
Various new physics models have been studied by the new approach of GW signals in the post aLIGO era~\cite{Huang:2016odd,Dev:2016feu,Jaeckel:2016jlh,Hashino:2016rvx,Jinno:2016knw,Yu:2016tar,Kobakhidze:2016mch,
Huang:2016cjm, Artymowski:2016tme, Soni:2016yes, Balazs:2016tbi, Dorsch:2016nrg, Huang:2017laj,
Chao:2017vrq,Beniwal:2017eik,Addazi:2017gpt,Marzola:2017jzl,Bian:2017wfv},
such as the GW detection of
the dark matter (DM)~\cite{Chao:2017vrq,Beniwal:2017eik}, 
the hidden sector~\cite{Dev:2016feu,Jaeckel:2016jlh, Soni:2016yes, Addazi:2017gpt,Huang:2017laj}
and the electroweak baryogenesis~\cite{Huang:2016odd,Huang:2016cjm,Artymowski:2016tme, Dorsch:2016nrg, Bian:2017wfv}. 


Another important puzzle in particle cosmology is the particle nature of the DM.
Usually there are three ways to search for  particle DM: direct detection, indirect detection, and collider searches.
With the experimental precision of the DM direct detection are gradually approaching the neutrino backgrounds, such as the recent LUX and PandaX-II experiments~\cite{Tan:2016zwf,Akerib:2016vxi},
GWs may become a new approach to explore the existence of DM since in a large classes of DM models
where a FOPT can be trigged by the DM particles and other associated particles~\cite{Espinosa:2008kw, Chowdhury:2011ga, Schwaller:2015tja}.
In general, a strong FOPT can be triggered by the DM candidates in generic new physics models, which can produce detectable GW spectrum by Evolved Laser Interferometer Space Antenna (eLISA)~\cite{Seoane:2013qna},  Big Bang Observer (BBO)~\cite{Corbin:2005ny}, Deci-hertz Interferometer Gravitational wave Observatory (DECIGO)~\cite{Seto:2001qf} and Ultimate-DECIGO (U-DECIGO)~\cite{Kudoh:2005as}.

In this paper, we study the  chance to explore  a general classes of DM models through phase transition GW signals
since the GW signal becomes a novel approach
to study the property of the various new physics models with extended Higgs sector
after the discovery of Higgs boson at LHC and GWs at aLIGO.
And the extended Higgs sector could incorporate the scalar DM candidate.
One generic classes of new physics models are the inert multiplet scalar DM models~\cite{Cirelli:2005uq,Hambye:2009pw} 
and mixed scalar DM models~\cite{Fischer:2013hwa,Cheung:2013dua}.
In these models, the scalar multiplets usually belong to a hidden sector, and thus
contain the DM candidate.
At the same time, we know that usually extra scalars could change the
phase transition structure for the standard model (SM) Higgs boson
around the  temperature of the electroweak scale.  Due to extra scalar degree of freedom in the thermal plasma,
the extra inert scalars could enhance the strength of the electroweak phase transition.
On the other hand, these models typically encounter the tension between tight direct detection constraints and the strong FOPT.
To avoid the tighter and tighter direct detection constraints, we could focus on the {\it blind spot} region, at which the Higgs-DM coupling is identically zero,
thus avoiding the DM spin-independent direct detection constraints.
In the scalar multiplet DM models, usually there is strong correlation between the Higgs-DM coupling and the strong FOPT: zero Higgs-DM coupling indicates there
is no effect on the electroweak phase transition from the inert DM sector.
The only exception in the single scalar multiplet DM models is the inert doublet model (IDM)~\cite{Ma:2006km, Barbieri:2006dq}, in which although the direct detection rate could be tiny
the strong FOPT could be realized.
%
%
In the blind spot region, requiring the observed relic abundance, the allowed parameter space in IDM is very limited~\cite{Belyaev:2016lok}.
To enlarge the allowed parameter region, we consider the mixed inert DM models, in which two inert scalars are mixed~\cite{Fischer:2013hwa,Cheung:2013dua},
and thus there is large  blind spot parameter region.
We will study the possibility to obtain strong FOPT in the blind spot parameter region in the mixed DM models.
We expect that the GW signature could further explore the parameter region which has not yet been explored by the
direct DM detection experiments.

This paper is organized as follows:
In Section II, we discuss the general inert scalar DM model, and
point out the IDM in the blind spot region could lead to the GW signals while avoiding the DM constraints.
In Section III, the mixed inert scalar models are investigated in the blind spot region with the DM constraints and the GW signals.
In Section IV, we show our final discussions and conclusions.

\section{Multiplet Dark Matters and Gravitational Waves}

Let us start from a class of DM models in which the SM is extended by adding an electroweak scalar multiplet.
The scalar multiplet is $Z_2$-odd under an imposed global $Z_2$ symmetry.
In certain hypercharge assignment, the neutral component in the scalar multiplet
could be the lightest particle in the multiplet, and thus it could be
a possible DM candidate.
This class of models have been well-studied in the inert isospin-singlet~\cite{Burgess:2000yq}, doublet~\cite{Ma:2006km, Barbieri:2006dq}, and triplet~\cite{FileviezPerez:2008bj} cases.
And a general DM multiplets in terms of the $SU(2)_L$ representations  have also been investigated systematically~\cite{Cirelli:2005uq, Hambye:2009pw}.
In general, the scalar multiplet $H_n$ belongs to the representation $n$ of the $SU(2)$ group, with hypercharge $Y$ under the transformation of the $SU(2)_L \times U(1)_Y$ gauge group
\bea
	H_n = \left( \begin{array}{c} h^{+j} \\ \vdots \\ h^{-j} \end{array}  \right) \sim (2j + 1, Y),
\eea
where the $T^3$ charge $j$ runs from $-\frac{n-1}{2}$ to  $\frac{n-1}{2}$.
The relevant Lagrangian is written as
\bea
	{\mathcal L} = D_\mu H_n^\dagger D^\mu H_n - V(\Phi, H_n),
\eea
where $\Phi$ is the SM Higgs doublet.
The scalar potential is usually taken to be
\bea
	V(\Phi, H_n) &=& V(\Phi) + M_n^2 H_n^2 + \lambda_2 H_n^4 + \lambda'_2 (H^\dagger_n \tau^{n}_i H_n)^2 \nn \\
	&+& \lambda_3 \Phi^2 H_n^2 + \lambda'_3 (\Phi^\dagger \tau^{2}_i \Phi)(H^\dagger_n \tau^{n}_i H_n),
	\label{eq:Vmulti}
\eea
where $V(\Phi) = -\mu^2 \Phi^2 + \lambda \Phi^4$.
For a real multiplet~\footnote{A real multiplet only has integer isospin. Thus there is no real doublet, quadruplet, etc.}, the term $H^\dagger_n \tau^{n}_i H_n$ is identical to zero, and thus the $\lambda'_2$ and $\lambda'_3$ terms disappear.
As a consequence, only the term $\lambda_3$ connects $H_n$ to $\Phi$, and the tree-level masses of the multiplet components are
degenerate with $m_n^2 = M_n^2 + \frac12 \lambda_3 v^2$~\footnote{The loop corrections to the tree-level masses cause small mass splitting between
charged components and neutral DM candidate. }.
Similarly, in the complex singlet model, the $\lambda'_2$ and $\lambda'_3$ terms do not exist.
For complex doublet model, there will be more terms than Eq.~(\ref{eq:Vmulti}) in the potential
\bea \label{vidm}
	V(\Phi, H_2) &=& V(\Phi) + M_D^2 H_2^2 + \lambda_2  H_2^4  \nn \\
	&+& \lambda_3 \Phi^2 H_2^2 + \lambda_4 |\Phi^\dagger H_2|^2\nn \\
	&+& \lambda_5/2 \left[(\Phi^\dagger H_2)^2 + h.c.\right].
\eea
To make the real component $H$ as the DM candidate, $\lambda_{345}=\lambda_3+\lambda_4+\lambda_5 < \lambda_3$
and $\lambda_5 < 0$ is needed.
For a complex multiplet with higher isospin, for simplicity we will only focus on the case with hypercharge $Y = 0$~\footnote{If the hypercharge $Y$ is not zero,
the neutral components of the complex multiplet have interaction with the $Z$ boson,
which usually leads to a large spin-independent cross-section between DM and the nucleon~\cite{Hambye:2009pw}. But if for some reason
there is a mass splitting between real and imaginary components of the neutral scalar, the direct detection constraints might be avoided.} in this work.
As shown in Ref.~\cite{Hambye:2009pw},  typically the complex triplet with $Y = 0$ can be viewed as
two real triplets.

Firstly, we discuss the FOPT induced by the inert scalar models.
Up to one-loop level, the effective potential at the finite temperature can be written as
\begin{equation}
V_{\mathrm{eff}}(\Phi,T) = V_{0}(\Phi)+V_{\rm CW}(\Phi) +
V_{\rm ther}(\Phi, T) +V_{\rm daisy}(\Phi, T),\nonumber
 \label{veff}
\end{equation}
where $V_{\rm ther}(\Phi, T) +V_{\rm daisy}(\Phi, T)$
is the thermal effects with the daisy resummation and
$V_{\rm CW}(\Phi)$ is the Coleman-Weinberg potential at zero temperature.
In the inert scalar models, the necessary potential barrier for the strong FOPT in the
effective potential at the finite temperature origins from thermal loop effects, where the bosons contribute to the  effective
potential with the form $V_{\rm T} \ni (-T/12 \pi) \bigl( m_{\rm
  boson}^2(\Phi,T) \bigr)^{3/2}$ in the limit of high-temperature expansion.
For the heavy fields whose masses are much larger than the critical temperature,
the contributions from heavy particles can be omitted due to Boltzmann suppression.
This can help to simplify our discussions when the models have many
new fields at different energy scales.
Thus, due to the above considerations and the fact that
we only study the inert scalar models and the thermal barrier induced strong FOPT,
the parameter space of very large mass scalar new bosons are not favored
since the field independent term in the thermal masses should be small enough
in order not to dilute the cubic terms.
We focus our study in the case without very heavy scalars.
Therefore, to begin with the concrete prediction of the GW signals in the following examples,
the general effective potential
near the phase transition temperature can be further approximated by
\begin{align}\label{vappro}
	V_{\rm eff}(h, T) \approx \frac{1}{2} \left( - \mu^2 + c \, T^2 \right) h^2 - \frac{ \varepsilon T}{12 \pi} h^3 + \frac{\lambda}{4} h^4.
\end{align}
Here, $h$ represents the Higgs boson field in the unitary gauge as $\langle \Phi \rangle= h/\sqrt{2}$, where the angle bracket means the vacuum expectation value (VEV) of the field.
The coefficient $\varepsilon$ quantifies the total interactions between the light bosons and the Higgs boson.
In the SM, we have $\varepsilon_{\rm SM} = \frac{6 m_W^3 + 3 m_Z^3}{v^3}$
and
$c_{\rm SM}=\frac{6 m_t^2+6 m_W^2+3m_Z^2+\frac{3}{2}m_H^2}{12 v^2}$.

Thus, in this case of qualitative analysis,
the washout parameter can be obtained as $\frac{\langle v \rangle (T_c)}{T_c} \approx \frac{\varepsilon}{6 \pi \lambda}$,
and it should be larger than one for a strong FOPT.
In inert scalar models, the washout parameter usually is approximately proportional to the Higgs-inert scalar coupling $\lambda_3$
for the potential in Eq.(\ref{eq:Vmulti}).
Roughly, a larger Higgs-inert scalar coupling produces a stronger FOPT.
For high multiplets, the situation becomes more complicated due to the large thermal mass coming from gauge interactions.
The large thermal masses on the new scalars cause plasma screening, and thus decrease the strength of the phase transition.
Therefore, there are two parameters which control the strong FOPT: the Higgs-inert scalar coupling,
and the thermal mass coming from gauge interactions.

We know that the Higgs-inert scalar coupling is usually constrained by the DM
direct detection experiments.
The tree-level interaction between the DM and the nucleon is through the Higgs boson exchange,
which induces the elastic spin-independent cross section
\bea \label{dm1}
	\sigma_{\rm SI} \simeq f_N^2 \frac{\lambda_{h\chi\chi}^2}{\pi} \left(\frac{m_N^2}{m_{\chi} m_h^2}\right)^2,
\eea
where the Higgs-DM coupling $\lambda_{h\chi\chi} = \lambda_{345}/2$ for inert doublet, and $\lambda_{h\chi\chi} = \lambda_3/2$ for other intert multiplet.
Current DM direct detection experiments~\cite{Tan:2016zwf,Akerib:2016vxi} constrain that the $\lambda_{h\chi\chi}$ is around $0.012$
for about 100 GeV DM mass.
In general, the coupling $\lambda_{h\chi\chi}$ in direct detection is the same as the one which mainly controls the strong FOPT.
Thus the tighter direct detection constraint, the less significance of the FOPT.
The only exception is the IDM.
In the IDM, due to the extra interaction terms in the scalar potential, the strong correlation between direct detection constraints and FOPT can be avoided.
We will focus on the blind spot region in IDM, in which although the direct detection constraint can be evaded, the  strong FOPT can be induced.
Thus, the corresponding phase transition GWs can be produced and the DM abundance
can be explained.

\subsection{Inert Singlet, Triplet and Multiplet Scalar Models}

In the inert singlet model, the  Higgs portal term $\lambda_3 |\Phi|^2 S^2/2$ is crucial,
where we use $S$ represent the inert singlet scalar instead of $H_1$.
The corresponding washout parameter in the inert singlet model is approximately
\begin{equation}\label{v1}
\frac{\langle v \rangle (T_c)}{T_c} \approx \frac{\frac{6 m_W^3 + 3 m_Z^3}{v^3}+ \left( \lambda_3/2\right)^{3/2}}{6 \pi \lambda}.
\end{equation}
In the inert triplet model,
we only consider a simple model with an $SU(2)_L^{}$ triplet scalar $H_3(1,3,0)$ with a zero hypercharge.
The relevant term involving the triplet scalar $H_3$ should be $\lambda_3 \Phi^\dagger_{}\Phi\textrm{Tr}(H_3^{2})$.
The corresponding washout parameter in the inert triplet model is approximately
\begin{equation}\label{v1}
\frac{\langle v \rangle (T_c)}{T_c} \approx \frac{\frac{6 m_W^3 + 3 m_Z^3}{v^3}+3 \left(\lambda_3 /2 \right)^{3/2}}{6 \pi \lambda}.
\end{equation}
However, in both models the direct DM search puts strong constraint  $\lambda_3 \lesssim 0.01$, which means the FOPT is forbidden by the DM
direct experiments.

For a general $n$-multiplet inert scalar $H_n$ models, there are two competing sources which affect the strong FOPT~\cite{AbdusSalam:2013eya}:
\bit
\item Higgs-inert multiplet coupling with $\varepsilon_{\rm new} \sim n \left( \lambda_3/2 \right)^{3/2}$;
\item plasma screening due to large thermal mass coming from gauge interactions.
\eit
Typically the higher multiplet, the larger screening and decoupling effects, which weaken the FOPT.
According to Ref.~\cite{AbdusSalam:2013eya}, for a multiplet with $n > 3$, the screening effects significantly decrease the strength of the
FOPT.
Furthermore, another severe constraint for high multiplet model is from Higgs  diphoton rate with all the charged scalars running in the loop. 
For the real scalar multiplet with $n > 3$,  the Higgs coupling measurement data put very strong constraints on the masses of the
scalar multiplet to be greater than 300 GeV, which makes the scalar degree of freedom decoupled from the plasma.

\subsection{Inert Doublet Model}

As mentioned above, usually the blind spot region with zero Higgs-DM coupling indicates the DM sector does not affect the
electroweak phase transition.
In IDM, zero Higgs-DM coupling does not indicate zero Higgs-inert scalar couplings are zero.
There is no correlation between direct detection and strong FOPT in IDM.
Therefore, we expect to obtain the strong FOPT and detectable GW signals in the blind spot parameter region.

We  investigate the finite temperature effective potential and  discuss conditions of strong FOPT
in detail~\cite{Borah:2012pu, Gil:2012ya, AbdusSalam:2013eya, Blinov:2015vma}.
The relevant scalar potential in the IDM is given in Eq.(\ref{vidm})
where $H_2$ stands for the inert doublet scalar without VEV. In the IDM, we assume that
only $\Phi$ can acquire VEV, namely $\Phi^T=(0,v+h)/\sqrt{2}$ and
H is the lightest component of the inert doublet $H_2$ with the mass $m_H^2=M_D^2 + \frac12 \lambda_{345}  v^2$.
Thus, the particle H is the DM candidate here.
The other neutral scalar mass is $m_A^2=M_D^2 + \frac12 (\lambda_3 + \lambda_4 - \lambda_5)  v^2$,
and the charged scalar mass is $m_{H^\pm}^2 = M_D^2 + \frac12 \lambda_3 v^2$.
The thermal phase transition with full 2-loop effective potential
has been studied recently~\cite{Laine:2017hdk} and it shows the
one-loop effective potential in the high temperature expansion is rather
reliable in the IDM. To clearly see the phase transition physics and simplify the following discussions on the phase transition
GW signals, we take the following approximation of the one-loop effective potential including the daisy resummation:
\begin{eqnarray}\label{ve_idm}
V_{\rm eff}(h, T) &  \approx & \frac{1}{2} \left( - \mu^2 + c_{IDM} \, T^2 \right) h^2 + \frac{\lambda}{4} h^4  \nonumber \\
 &-&   \frac{T}{12 \pi} \Sigma n_b (m_b^2 (h,T))^{3/2}  \nonumber \\
 &-& \Sigma n_b \frac{m_b^4(h,T)}{64 \pi^2} \log\frac{m_b^2(h,T)}{c_a T^2}                           \nonumber \\
 &-&  n_t \frac{m_t^4(h)}{64 \pi^2} \log\frac{m_t^2(h)}{c_b T^2},
\end{eqnarray}
where $\log c_a =5.408$ and $\log c_b=2.635$.
In the effective potential, the particles running in the loop are
the particles in the model
with the following degrees of freedom:
\bea
n_{W^{\pm}}=4, \  n_{Z}=2, \  n_{\pi}=3, \nonumber \\
\  n_h=n_{H} = n_{H^+}=n_{H^+} = 1, \  n_{t}=-12. \nonumber
\eea
The field-dependent masses of the gauge bosons and the top quark at zero temperature are given by
\bea
m_{W}^2(h) &=& \frac{g^2}{4} h^2, \
m_{Z}^2(h) = \frac{g^2+g'^2}{4} h^2,
m_{t}^2(h) = \frac{y_t^2}{2} h^2, \nn
\label{masses}
\eea
where $y_t$ is the top Yukawa coupling.
The field-dependent thermal masses at the temperature $T$ are
\bea
    m_{h}^2(h, T) &=&m_{\pi}^2 \approx 3\lambda h^2 - \mu^2  + c_1 T^2, \nonumber \\
	m_H^2(h, T)&\approx&  \frac12 (\lambda_3 + \lambda_4 + \lambda_5)h^2  + M_D^2+ c_2 T^2,\nonumber\\
	m_A^2(h, T)&\approx& \frac12 (\lambda_3 + \lambda_4 - \lambda_5)h^2   + M_D^2+ c_2 T^2,\nonumber\\
	m_{H^\pm}^2(h, T) &\approx&  \frac12 \lambda_3 h^2                    + M_D^2+ c_2 T^2, \nonumber
\eea
where $c_1 = \frac{\lambda}{2} +\frac{2\lambda_{3}+\lambda_4}{12} + \frac{3g^2 + g^{'2}}{16}
	+ \frac{y_t^2}{4}$ and
$c_2= \frac{\lambda_2}{2} + \frac{2\lambda_{3}+\lambda_4}{12} + \frac{3g^2 + g^{'2}}{16}.$
In the above formulae, we have considered the contribution from daisy resummation, which
reads as
\bea
	V & \supset & - \frac{T}{12\pi} \sum_{i = {\rm b}} n_b\left(\left[m_i^{2}(h,T)\right]^{3/2} -  \left[m_i^{2}(h)\right]^{3/2} \right).\nonumber
\eea
Here, the thermal field-dependent masses
$m_i^2(h, T) \equiv m_i^2(h)+ \Pi_i(h, T)$,
where $\Pi_i (h, T)$ is the bosonic field $i$'s self-energy in the IR limit.
%
%
%
As in the SM, this cubic term is the unique source to produce a thermal barrier  in the effective potential, and in the Higgs sector extended  models,
the new degree of freedoms in the inert scalar models  increase the barrier
and hence produce strong FOPT.
However, the cubic terms should be large enough to produce a strong FOPT.
To avoid diluting the cubic contribution to the thermal barrier, the Higgs boson field independent term needs to be
very small~\cite{Chowdhury:2011ga,Chung:2012vg}. Thus, in this limit,
we have $\varepsilon_{\rm IDM} \approx \varepsilon_{\rm SM} + 2 \left( \frac{\lambda_3}{2} \right)^{3/2} + \left( \frac{ \lambda_3 + \lambda_4 - \lambda_5}{2} \right)^{3/2} + \left( \frac{\lambda_{345}}{2} \right)^{3/2}$
and
$c_{\rm IDM} \approx c_{\rm SM} + \frac{2\lambda_3+\lambda_4}{12}$.
Then, the corresponding washout parameter in the IDM is roughly
\begin{equation}\label{v1}
\frac{\langle v \rangle (T_c)}{T_c} \approx \frac{e_{\rm SM}+2 \left( \frac{\lambda_3}{2} \right)^{3/2} + \left( \frac{ \lambda_3 + \lambda_4 - \lambda_5}{2} \right)^{3/2} + \left( \frac{\lambda_{345}}{2} \right)^{3/2}}{6 \pi \lambda}. \nonumber
\end{equation}

From Eq.(\ref{dm1}) and the washout parameter, we find that small DM direct detection rate and strong FOPT can be realized when we take the
blind spot region, in which $\lambda_{h\chi\chi}=\lambda_{345}/2$  approaches to zero~\footnote{ $\lambda_{345}$ can be very small due to the cancellation between three couplings $\lambda_3$, $\lambda_4$ and $\lambda_5$ while keeping $\lambda_3$ large enough to produce a strong FOPT.}.
The DM model of IDM needs to satisfy the required DM relic density observed from Planck: $\Omega h^2 = 0.1184 \pm 0.0012$~\cite{Ade:2015xua}.
This put very strict constraints on the IDM: the DM mass is determined to be $m_{\chi} > 540$ GeV~\cite{Hambye:2009pw} except for
the parameter region with large mass splitting between charged and neutral components.
For the DM mass lower than $m_h/2$, the Higgs invisible decay puts very tight constraint on the parameter space.
According to the latest study~\cite{Belyaev:2016lok}, there are two viable mass regions:
\bit
\item near Higgs funnel region with large mass splitting between charged and neutral components: $m_\chi$ around $55 \sim 75$ GeV with $\lambda_{345} < 0.04$;
\item heavy DM region: $m_H > 540$ GeV with $\lambda_{345}$ in a broader range as $m_\chi$ gets heavier.
\eit
To keep the scalar non-decoupled from the thermal plasma, it is necessary to have light DM.
The dominant DM annihilation channel will be $\chi\chi \to WW^*,ZZ^*$ with contact, $t$- and $s$-channels.
We will focus on the DM mass around $55 \sim 75$ GeV, and the blind spot region with $\lambda_{345} \simeq 0$. 
%
%
Combined the direct DM constraints, the DM relic density, collider constraints~\cite{Belyaev:2016lok} and the conditions
for strong FOPT, this light mass region $55 \sim 75$ GeV is favored.
The strong FOPT can be produced
if $\lambda_3/2$ and $(\lambda_3+\lambda_4-\lambda_5)/2$ are order $1$, then detectable GW signals can be produced,
while keep the coupling between Higgs boson and DM pair small enough to satisfy DM direct experiments and relic
density.

\begin{figure}
\begin{center}
\includegraphics[scale=0.45]{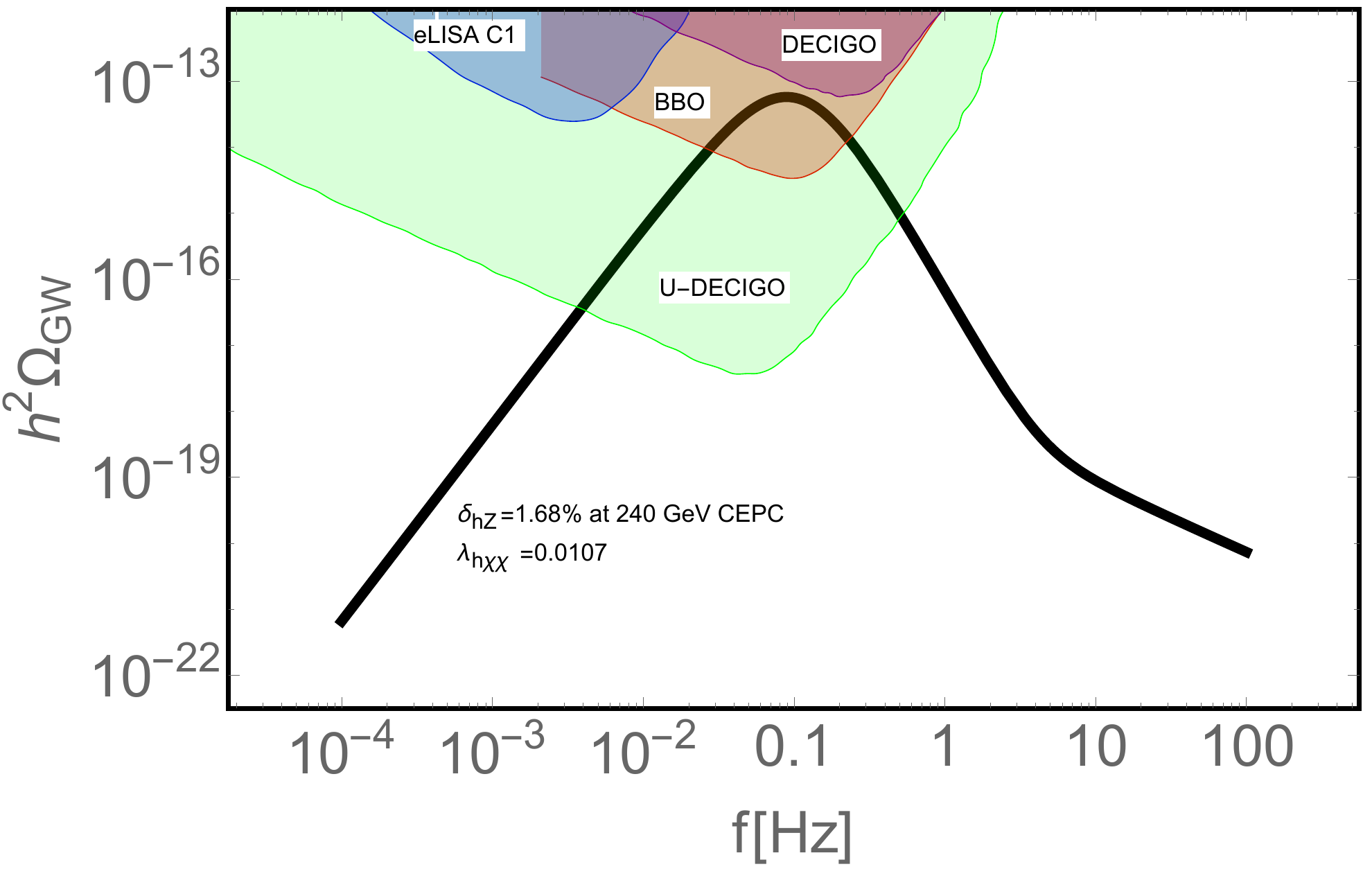}
\caption{The phase transition GW spectra $h^2\Omega_{\rm GW}$ in the IDM. The colored regions  represent the expected sensitivities of GW interferometers U-DECIGO, DECIGO, BBO and eLISA, respectively. The black line depicts the GW spectra in the IDM for the set of benchmark points, which also represents
the corresponding hZ cross section deviation at the 240 GeV CEPC and the corresponding DM coupling.}
\label{idm2}
\end{center}
\end{figure}

Considering the above discussion, we take one set of benchmark points $\lambda_3=2.84726$, $\lambda_4=\lambda_5=-1.41293$ and $M_D=60.89\rm~GeV$.
Then, the corresponding DM mass is 66 GeV, the pseudo scalar mass and the charged scalar mass are both 300 GeV,
$\lambda_{h\chi\chi} =\lambda_{345}/2=0.0107$.
Taking this set of benchmark points, the relic density, DM direct search, collider constraints and a strong FOPT can be satisfied simultaneously.
Using the methods and formulae in the appendix, the phase transition GW signal is shown in Fig.\ref{idm2}, which is just
within the sensitivity of BBO and U-DICIGO.
The colored regions represent the sensitivities of different GW experiments (DECIGO~\cite{Moore:2014lga}, eLISA~\cite{Caprini:2015zlo}, BBO, and U-DECIGO~\cite{Kudoh:2005as}), and the black line corresponds to the GW signals, which also means
the hZ cross section ($e^+ + e^- \to h + Z$) deviation from the SM in 240 GeV circular electron-positron collider (CEPC).
At the 240 GeV CEPC~\cite{CEPC-SPPCStudyGroup:2015csa} with an integrated luminosity of $10~\mathrm{ab}^{-1}$, the precision of $\sigma_{hZ}$ could be about $0.4\%$~\cite{Gomez-Ceballos:2013zzn}.
And at the 240 GeV CEPC, the deviation of the hZ cross section $\delta_{hZ}\equiv \frac{\sigma-\sigma_{SM}}{\sigma_{SM}}$ at one-loop level~\cite{Huang:2015izx} is
about $1.68\%$~\cite{Arhrib:2015hoa,Kanemura:2016sos}, which is well within the sensitivity of CEPC.
The international linear collider (ILC)~\cite{Gomez-Ceballos:2013zzn} can also test this model.
The GW signal and the hZ cross section deviation at future lepton collider can
make a double test on the DM of IDM as shown in Fig.~\ref{idm2}.

\section{Gravitational Waves in Mixed Dark Matters}

We have discussed FOPT and GWs if there is only one single multiplet scalar dark matter in the dark sector.
Due to the tight correlation between strong FOPT and
the DM direct detection, only the IDM is viable for strong FOPT and detectable GWs.
Based on the relic density requirement, the IDM has very limited viable parameter region: $m_\chi \simeq 55 \sim 75$ GeV, and the blind spot region $\lambda_{345} \simeq 0$.
In this section, we would like to extend the single DM multiplet models into the mixed scalar DM models.

The mixed scalar DM scenario involves in several $Z_2$-odd scalar multiplets in the dark sector, which could be mixed.
The simplest models involve in two dark matter multiplets: the mixed   singlet-doublet model (MSDM) and the mixed   singlet-triplet model (MSTM)~\cite{Fischer:2013hwa,Cheung:2013dua}.
%
Compared to the single scalar DM, the advantages of the mixed DM scenario are
\bit
	\item It is easy to obtain a large blind spot region, at which the DM-Higgs coupling is zero.; 
	\item If the DM is mixture of a singlet and multiplet (well-tempered), the relic density could be realized in a large DM mass range;
	\item There are more degree of freedoms which contribute to the thermal barriers of the electroweak phase transition.
\eit
From the above, we note that it is necessary to have a singlet component to reduce the large annihilation cross section during freeze-out.
The higher multiplet $n >3$ is also tightly constrained by the plasma screening and Higgs diphoton constraints.
%



We will focus on the strong FOPT and GWs in the blind spot region with broad DM mass range in the MSDM and  MSTM.
Here, we adapt the effective Lagrangian in Ref.~\cite{Cheung:2013dua}.
Denote the SM singlet as $S$ and the neutral component of the multiplet as $H_n^0$.
Due to the mixing between two fields, the mass eigenstates have
\bea
	\chi &=& \cos \theta S - \sin \theta H_n^0, \\
	 s   &=& \sin \theta S + \cos \theta H_n^0.
\eea
The effective Lagrangian for the mass eigenstates has
\bea
	{\mathcal L}_{\rm eff} = a_{h\chi\chi} h \chi\chi + a_{h ss} h ss + \cdots.
\eea
Define the dimensionless coupling $\lambda_{h\chi\chi}=a_{h\chi\chi}/(2v)$ and $\lambda_{hss}=a_{hss}/(2v)$.
According to Ref.~\cite{Cheung:2013dua}, the mixed DM scenario could have large viable parameter region
after imposing the direct detection constraints and requirement on the DM relic abundance.
Unlike the IDM, the DM mass could be arbitrary.
The viable parameter region strongly  depends on two parameters: the $\lambda_{h\chi\chi}$, and the mixing angle $\theta$.
The spin-independent direct detection rate is dominated by the Higgs exchange with the cross section: 
\bea
	\sigma_{\rm SI} \simeq f_N^2 \frac{\lambda_{h\chi\chi}^2}{\pi} \left(\frac{m_N^2}{m_{\chi} m_h^2}\right)^2.
\eea
Thus, the $\lambda_{h\chi\chi}$ needs to be small to avoid tight direct detection constraints.
In this study, we take the blind spot region: $\lambda_{h\chi\chi} \sim 0$.
The dominant thermal annihilation channels could be very different and distinct depending on the DM mass region.
To produce detectable GW signature, we would like to focus on the moderate DM mass region ($m_W < m_\chi < 300$ GeV)~\footnote{Of course, the parameter space near the Higgs funnel region is also viable.
But we focus on a broader parameter region in this work.}.
The dominant annihilation channels are
\bit
\item the $s$-channel annihilations: $\chi \chi \to h \to WW, ZZ, ff$, in which the annihilation rate depends on the $\lambda_{h\chi\chi}$;
\item the contact four point annihilation: $\chi \chi \to   WW$ for real multiplet and $\chi \chi \to   WW, ZZ$ for doublet;
\item the $t$-channel annihilation: $\chi \chi \to   WW$ via exchanging of the charged scalars;
\eit
Given   $\lambda_{h\chi\chi} \sim 0 $ in the blind spot region, we assume that the contact and $t$-channel annihilation processes are dominant~\footnote{
There is still small parameter region in which the $s$-channel and gauge interactions have interference effects.}.
In this case, the annihilation cross sections only depend on the gauge interactions, and the annihilation rate near the blind spot region is
greatly simplified as
\bea
\sigma_{\chi \chi \to   VV} v^2 \simeq \frac{ 3 g^4 + 6 g^2 g'^2 + g'^2}{256 \pi  m_\chi^2}\sin^2\theta,
\eea
for the mixed singlet-doublet, and
\bea
	\sigma_{\chi \chi \to   WW} v^2 \simeq \frac{g^4}{4 \pi m_\chi^2}\sin^2\theta,
\eea
for the mixed singlet-triplet model.
To obtain the required relic abundance, the mixing angle $\sin\theta$ needs to satisfy
\bea
\sin\theta & \simeq&\frac{m_\chi}{540~ \text{GeV}} \,\,{\rm  (doublet)}, \\
\sin\theta & \simeq&\frac{m_\chi}{2000~ \text{GeV}} \,\,{\rm  (triplet)},
\label{eq:mixing}
\eea
which could be deduced from Ref.~\cite{Cirelli:2005uq, Hambye:2009pw}.
Note that this simple relation is only valid near the blind spot region.
In the interested mass region ($m_W < m_\chi < 300$ GeV), the well-tempered DM is favored.
%
%
Therefore, we could choose the benchmark points 
by acquiring  $\lambda_{h\chi\chi}$ is close to zero (blind spots) and $\sin\theta$ is determined by the above Eq.~(\ref{eq:mixing}).
Using these two conditions, we could obtain correct relic abundance, while escaping the DM direct detection constraints and obtaining a strong FOPT with detectable GWs.

\subsection{The Mixed Singlet-Doublet Model}

The   MSDM contains a singlet $S$ with hypercharge $Y = 0$, and a doublet with hypercharge $Y = 1/2$.
The tree level potential in the  MSDM is
\begin{eqnarray}
V_0 & = & \frac{1}{2} M_S^2 S^2
+ M_D^2 H_2^{\dagger}H_2 + \frac{1}{2} \lambda_S S^2 \left|\Phi\right|^2
+ \lambda_3 \Phi^{\dagger} \Phi H_2^{\dagger}H_2 \nonumber \\
&+&  \lambda_4 |{\Phi^{\dagger}H_2}|^2 + \frac{\lambda_5}{2} [(\Phi^{\dagger}H_2)^2 + h.c. ] +
A \left[ S \Phi H_2^\dagger + h.c. \right]\,.\nn
\end{eqnarray}
Here, we omit all interactions involving only $S$ and $H_2$ and the kinetic terms.
The mixed mass matrices are
\bea
{\mathcal M}_{\rm neutral}^2 =
\left(\begin{array}{cc}
\tilde{M}_S^2  & A v \\
A v &  \tilde{M}_D^2
\end{array}
\right),
\eea
where $\tilde{M}_S^2 = M_S^2 + \frac{1}{2} v^2 \lambda_S $ and
$\tilde{M}_D^2 = M_D^2 + \frac{1}{2} v^2 \lambda_{345}$.
The two eigenvalues of the mass-squared
matrix are
\begin{eqnarray}
\frac{1}{2} \left[ \tilde{M}_S^2 + \tilde{M}_D^2 \pm \sqrt{\left(
    \tilde{M}_S^2 - \tilde{M}_D^2 \right)^2 + 4 v^2 A^2} \right]\,,
\end{eqnarray}
with the smaller eigenvalue corresponding to DM mass square
\begin{equation}
 m_\chi^2=\frac{1}{2} \left[ \tilde{M}_S^2 + \tilde{M}_D^2 - \sqrt{\left(
    \tilde{M}_S^2 - \tilde{M}_D^2 \right)^2 + 4 v^2 A^2} \right].
\end{equation}
The mixing angle is
\bea
	\tan2\theta = \frac{2A v}{\tilde{M}_S^2 - \tilde{M}_D^2}.
\eea
The coupling between the Higgs boson and DM is
\begin{eqnarray}
a_{h\chi\chi} & = & \frac{1}{2} v \left( \lambda_S + \lambda_{345} \right) \\
&-& \frac{2 v A^2 + \frac{1}{2}v \left(\tilde{M}_S^2
  - \tilde{M}_D^2 \right) \left(\lambda_S - \lambda_{345} \right)}{\sqrt{\left(\tilde{M}_S^2 - \tilde{M}_D^2
    \right)^2 + 4 v^2 A^2}} \,,\nn
\label{hxx}
\end{eqnarray}
and the coupling between the Higgs boson and the
singlet scalar is
\begin{eqnarray}
a_{hss} & = & \frac{1}{2} v \left( \lambda_S + \lambda_{345} \right) \\
&+&  \frac{2 v A^2 + \frac{1}{2}v \left(\tilde{M}_S^2
  - \tilde{M}_D^2 \right) \left(\lambda_S - \lambda_{345} \right)}{\sqrt{\left(\tilde{M}_S^2 - \tilde{M}_D^2
    \right)^2 + 4 v^2 A^2}} \,.\nn
\label{hss}
\end{eqnarray}


In the DM favored parameter spaces (small $\lambda_{h\chi\chi}$, $sin\theta \approx m_\chi/540$ GeV and $m_W<m_\chi<300$ GeV),
a strong FOPT can be induced due to the fact that the strong FOPT can occur in the IDM and the mixing with an inert singlet can
enhance the FOPT. The mixing term $A \left[ S \Phi D^\dagger + h.c. \right]$ does not significantly change the phase transition property
except for the change of the eigenvalues of the masses.
Taking the same approximations as in the IDM and writing in the forms of Eq.(\ref{vappro}), we have
$\varepsilon_{\rm MSDM} \approx \varepsilon_{\rm SM} + 2 \left( \frac{\lambda_3}{2} \right)^{3/2} + \left( \frac{ \lambda_3 + \lambda_4 - \lambda_5}{2} \right)^{3/2} + \left( \lambda_{h\chi\chi} \right)^{3/2}+ \left( \lambda_{hss} \right)^{3/2}$.
The strong FOPT can be easily realized if $\lambda_3$ and $\lambda_{hss}$ are order one which are allowed by the above DM constraints.

\begin{figure}
\begin{center}
\includegraphics[scale=0.46]{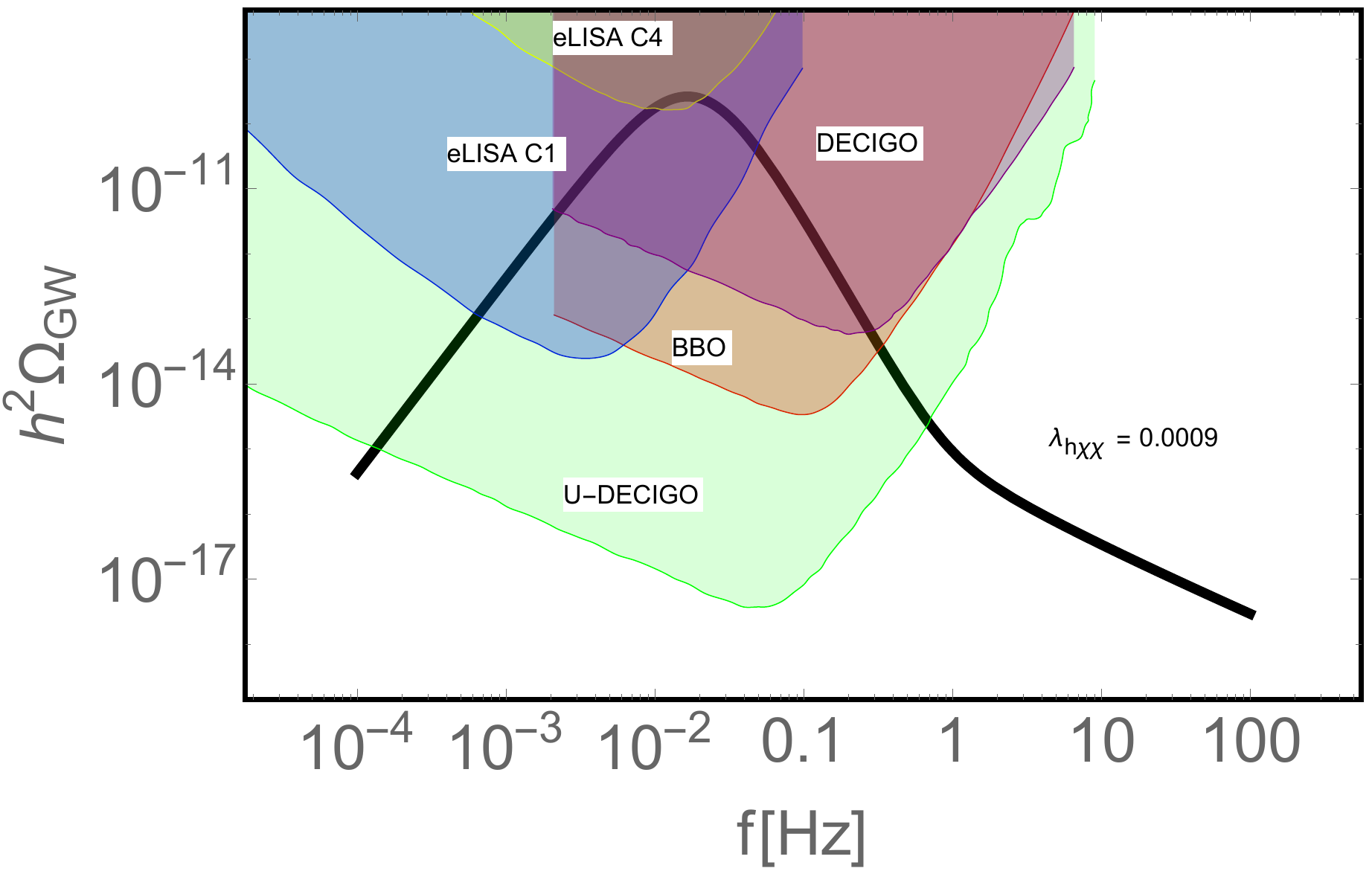}
\caption{The phase transition GW spectra $h^2\Omega_{\rm GW}$ in the MSDM. The colored regions  represent the expected sensitivities of GW interferometers U-DECIGO, DECIGO, BBO and eLISA, respectively. The black line depicts the GW spectra in the MSDM and the corresponding DM coupling for the set of benchmark points.}
\label{mix1}
\end{center}
\end{figure}

We take one set of benchmark points $\lambda_3=3.006$, $\lambda_4=-1.5$, $\lambda_5=-1.5$, $\lambda_S=4.006$, $A=91~\text{GeV}$ and
$M_D=M_S=117.5\rm~GeV$. Then, we have $m_\chi=100.6\rm~ GeV$, $\lambda_{h \chi\chi}=0.0009$ and $\lambda_{hss}=2.005$.
The corresponding GW spectra are shown in Fig.\ref{mix1}.
This set of benchmark points can evade the constraints from  DM direct experiments, give the correct relic density and
produce detectable GW signals by eLISA, BBO, DECIGO and U-DECIGO.
We estimate the deviation of the hZ cross section $\delta_{hZ}$ at one-loop level is
about $2.78\%$~\cite{Arhrib:2015hoa,Kanemura:2016sos} at the 240 GeV CEPC, which is well within the sensitivity of CEPC and ILC.
As for the precise prediction on the collider signals at the CEPC and ILC in this MSDM model, we leave it in our future study.


\subsection{The Mixed Singlet-Triplet Model}

\begin{figure}
\begin{center}
\includegraphics[scale=0.41]{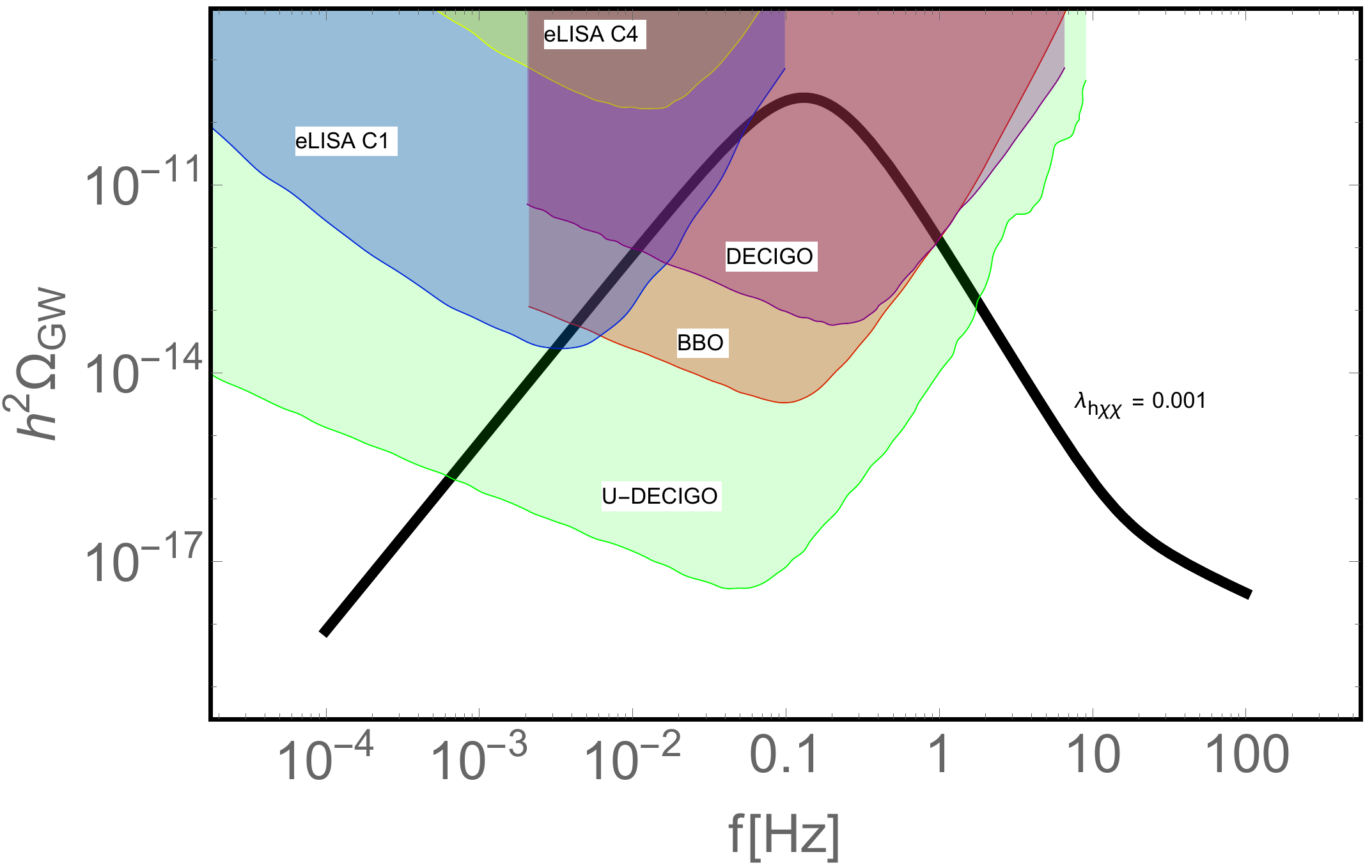}
\caption{The phase transition GW spectra $h^2\Omega_{\rm GW}$ in the MSTM. The colored regions  represent the expected sensitivities of GW interferometers U-DECIGO, DECIGO, BBO and eLISA, respectively. The black line depicts the GW spectra and
the corresponding DM coupling in the MSTM for the set of benchmark points.}
\label{mixt1}
\end{center}
\end{figure}

Similarly, the  MSTM contains a real singlet and a real triplet with $Y = 0$.
The relevant potential in the  MSTM is
\bea
	V_0 & = & \frac{1}{2} M_S^2 S^2 + M_{\Sigma}^2\textrm{Tr}(H_3^2)
+\kappa_\Sigma^{}\Phi^\dagger_{}\Phi\textrm{Tr}(H_3^{2}) \nonumber \\
&+& \frac{\kappa}{2} |\Phi|^2 S^2+\xi S \Phi^{\dagger}H_3 \Phi.
\eea
The coupling between the Higgs boson and DM is
\begin{eqnarray}
a_{h\chi\chi} & = & \frac{1}{2} v \left(\kappa +
\kappa_{\Sigma} \right) \\
&-& \frac{\frac{1}{4} v^3 \xi^2 + \frac{1}{2}v \left(\tilde{M}_S^2
  - \tilde{M}_{\Sigma}^2 \right) \left(\kappa -
\kappa_{\Sigma}  \right)}{\sqrt{\left(\tilde{M}_S^2 - \tilde{M}_{\Sigma}^2
    \right)^2 + \frac{1}{4} v^4 \xi^2}} \,,\nn
\label{thxx}
\end{eqnarray}
with the DM mass square
\begin{equation}
 m_\chi^2=\frac{1}{2} \left[ \tilde{M}_S^2 + \tilde{M}_{\Sigma}^2 - \sqrt{\left(
    \tilde{M}_S^2 - \tilde{M}_{\Sigma}^2 \right)^2 + \frac{\xi^2}{4} v^4} \right],
\end{equation}
where $\tilde{M}_S^2 = M_S^2 + \frac{1}{2} v^2 \kappa $ and
$\tilde{M}_{\Sigma}^2 = M_{\Sigma}^2 + \frac{\kappa_{\Sigma}}{2} v^2$.
The coupling between the Higgs boson and the singlet is
\begin{eqnarray}
a_{hss} & = & \frac{1}{2} v \left(\kappa +
\kappa_{\Sigma} \right) \\
&+& \frac{\frac{1}{4} v^3 \xi^2 + \frac{1}{2}v \left(\tilde{M}_S^2
  - \tilde{M}_{\Sigma}^2 \right) \left(\kappa -
\kappa_{\Sigma}  \right)}{\sqrt{\left(\tilde{M}_S^2 - \tilde{M}_{\Sigma}^2
    \right)^2 + \frac{1}{4} v^4 \xi^2}} \,.\nn
\label{thss}
\end{eqnarray}
Taking the same approximations as in the above discussions, we have
$\varepsilon_{\rm MSTM}=\varepsilon_{\rm SM} + \left( \lambda_{h\chi\chi} \right)^{3/2}+ ( \lambda_{hss})^{3/2}+2 \left( \kappa_\Sigma/2 \right)^{3/2}$.
In the MSTM, we take the set of benchmark points $\kappa=0.01,~\kappa_{\Sigma}=3.0$, $\xi=0.31337$, $M_\Sigma=50$ GeV, $M_S=119.93\rm~GeV$. Then, we have
$m_\chi=120\rm ~GeV$, $\lambda_{h\chi\chi}=0.001$ and $\lambda_{hss}=1.504$.
The GW spectrum is shown in Fig.\ref{mixt1}, which is within the sensitivity of
the BBO, eLISA, DECIGO and U-DECIGO. We also leave the study of collider signals in the MSTM in our future study.

\section{Conclusion} 

We have investigated the GW signatures associated with the possible strong
FOPT in the early universe in a generic  class of inert scalar multiplet DM models.
We have found that there is usually a strong tension between the strength of the phase transition
and the DM direct detection, except the blind spot region with narrow DM mass range in the IDM.
In the IDM, we have shown that although the
direct detection experiments could not probe the blind spot region, at which the Higgs-DM coupling is zero, the
GW detection and lepton colliders could help us further explore the
parameter space which cannot be addressed by the DM searches.

To enlarge the blind spot region with broader DM mass range, we considered the  mixed scalar DM scenario.
We focused on the minimal singlet-doublet dark matter and minimal singlet-triplet DM models,
and showed that the blind spot region in these models could avoid all the relevant constraints,
have required relic density, and produce strong FOPT in the
early universe.
%
%
The GW signatures are explored in this scenario, and we found that
the  blind spot region could be further explored via the GW signatures and collider signatures in future.
This provides us a new way to detect the DM in future.
And our strategy could also be used to more general DM scenarios than the scalar multiplet DM models.
%
We expect the future GW and collider experiments, such as eLISA and TianQin~\cite{Luo:2015ght}, CEPC and ILC,
could probe these DM scenarios beyond the reaches of the DM direct detection experiments.

\section*{Acknowledgements}

JHY was supported in part under U.S. Department of Energy contract DE-SC0011095.
FPH was supported in part by the NSFC (Grant Nos. 11121092, 11033005, 11375202) and by the CAS pilotB program.
FPH was also supported by the China Postdoctoral Science Foundation under Grant No. 2016M590133.
\appendix

\section*{Appendix}
In this appendix, the details on the calculation of phase transition GW spectrum are shown.
In general, there exist three mechanisms to produced phase transition GWs:
the first one is the well-known bubble collisions~\cite{Kamionkowski:1993fg},
the second one is the turbulence in the fluid, where a certain fraction of the bubble walls energy is converted into turbulence~\cite{Kosowsky:2001xp, Caprini:2009yp},
and the last one is the new mechanism of sound waves~\cite{Hindmarsh:2013xza}.

When a strong FOPT occurs, via thermally fluctuating or quantum tunneling the potential barrier, bubbles are nucleated with the
nucleation rate per unit volume $\Gamma = \Gamma_0(T) {\mathrm e}^{-S_E(T)}$ and $\Gamma_0(T)\propto T^4$~\cite{Linde:1981zj}.
The Euclidean action $S_E(T)\simeq S_3(T)/T$~\cite{Coleman:1977py,Callan:1977pt}, and then
$\Gamma = \Gamma_0 {\mathrm e}^{-S_3 / T}$~\cite{Linde:1981zj} where
\begin{align}
S_3(T)
&= \int d^3x \left[ \frac{1}{2} (\nabla \Phi)^2 + V_{\mathrm{eff}}(\Phi,T) \right].
\label{eq_bounceS3}
\end{align}
Thus, the bubble nucleation rate can be obtained~\cite{Cai:2017cbj} by calculating the  profile of the scalar field  $\Phi$
through solving the following bounce equation:
\begin{align}
\frac{d^2 \Phi}{dr^2} + \frac{2}{r} \frac{d\Phi}{dr} - \frac{\partial V_{\mathrm{eff}}(\Phi,T)}{\partial \Phi}
&= 0,
\end{align}
with the boundary conditions $\frac{d\Phi}{dr} (r=0)
= 0$ and $\Phi(r = \infty)
= 0$.
The phase transition terminates if it satisfies
\begin{equation}\label{tn}
S_3(T_{\ast})/T_{\ast} =4\ln (T_{\ast}/100 \mbox{GeV})+137.
\end{equation}

The phase transition GW spectrum depends on four parameters.
The first parameter is $\alpha \equiv \frac{\epsilon(T_{\ast})}{\rho_{\rm rad}(T_{\ast})}$,
where $T_{\ast}$ is
determined by Eq.(\ref{tn}).
The false vacuum energy (latent heat) density $\epsilon(T_{\ast}) = [T \frac{dV_{\rm eff}^{\rm min}}{dT} -V_{\rm eff}^{\rm min}(T) ]|_{T=T_{\ast}}$ , and the plasma thermal energy density $\rho_{\rm rad}(T_{\ast})=\frac{\pi^2}{30} g_{*}(T)T^4$.
The parameter $\alpha$ represents the strength the FOPT, namely, a larger value of $\alpha$ produces stronger GW signature.
The second parameter is $\frac{\beta}{H_*}
= \left. T \frac{d(S_3 / T)}{dT} \right|_{T=T_*}$, where one has
$\beta \equiv -\frac{d S_E}{d t} |_{t=t_{\ast}} \simeq\frac{1}{\Gamma}\frac{d \Gamma}{d t} |_{t=t_{\ast}}$, namely,
$\beta^{-1}$ corresponds to the typical time scale of the phase transition.
The third parameter is the efficiency factor $\lambda_i$ (i=co,tu,sw), and the last parameter is the bubble wall velocity $v_{b}$.

Once the four parameters are known, the corresponding phase transition GWs in the three mechanisms
can be directly obtained after considering the red-shift effects $\frac{a_\ast}{a_0}= 1.65 \times 10^{-5} \mbox{Hz}\times\frac{1}{H_{\ast}} \Big( \frac{T_{\ast}}{100 \rm GeV} \Big) \Big( \frac{g^t_\ast}{100} \Big)^{1/6}$.
The current peak frequency at each mechanism is $f_{\rm i}=f_{\rm i}^\ast a_\ast/a_0$ with i=co,tu,sw, respectively.
In the bubble collision mechanism, the GW spectrum is  expressed as~\cite{Huber:2008hg}
\begin{align}
 \Omega_{\rm co} (f) h^2 \simeq
 &1.67\times 10^{-5} \Big( \frac{H_{\ast}}{\beta} \Big)^2 \Big( \frac{\lambda_{co} \alpha}{1+\alpha} \Big)^2 \Big( \frac{100}{g^t_\ast} \Big)^{\frac{1}{3}} \nonumber \\
 &\times \Big( \frac{0.11v_b^3}{0.42+v_b^3} \Big) \Big[ \frac{3.8(f/f_{\rm co})^{2.8}}{1+2.8(f/f_{\rm co})^{3.8}} \Big] \nonumber
\end{align}
with the peak frequency $f_{\rm co}^\ast=0.62\beta/(1.8-0.1v_{b} +v_{b}^{2})$~\cite{Huber:2008hg} at $T_{\ast}$.
In the turbulence mechanism, the GW signals have the peak frequency at about $f_{\rm tu}^\ast=1.75\beta/v_b$ at $T_{\ast}$~\cite{Caprini:2015zlo},
and the phase transition GW intensity is formulated by \cite{Caprini:2009yp, Binetruy:2012ze}
\begin{align}
\Omega_{\rm tu} (f) h^2\simeq \nonumber
& 3.35\times 10^{-4}\Big(\frac{H_{\ast}}{\beta}\Big) \Big(\frac{\lambda_{\rm tu} \alpha}{1+\alpha}\Big)^{3/2}
\Big(\frac{100}{g^t_\ast}\Big)^{\frac{1}{3}}v_b\\
&\times\frac{(f/f_{\rm tu})^3}{(1+f/f_{\rm tu})^{11/3}(1+8\pi fa_0/(a_\ast H_\ast))}. \nonumber
\end{align}
In the sound wave mechanism, the GW spectrum can be written
as~\cite{Hindmarsh:2013xza, Caprini:2015zlo}
\begin{align}
\Omega_{\rm sw} (f) h^2\simeq \nonumber
& 2.65\times 10^{-6}\Big(\frac{H_{\ast}}{\beta}\Big) \Big(\frac{\lambda_{sw} \alpha}{1+\alpha}\Big)^2
\Big(\frac{100}{g^t_\ast}\Big)^{\frac{1}{3}}v_b\\
&\times\Big[\frac{7(f/f_{\rm sw})^{6/7}}{4+3(f/f_{\rm sw})^2}\Big]^{7/2}\nonumber
\end{align}
with the peak frequency  $f_{\rm sw}^\ast=2\beta/({\sqrt{3}v_b})$ at $T_{\ast}$~\cite{Hindmarsh:2013xza, Caprini:2015zlo},
where $\lambda_{sw}\simeq \alpha \left(0.73+0.083\sqrt{\alpha}+\alpha\right)^{-1}$ for relativistic bubbles~\cite{Espinosa:2010hh}.

\end{document}